\documentclass{article}

\usepackage{arxiv}

\usepackage[utf8]{inputenc} 
\usepackage[T1]{fontenc}    
\usepackage{hyperref}       
\usepackage{url}            
\usepackage{booktabs}       
\usepackage{amsfonts}       
\usepackage{nicefrac}       
\usepackage{microtype}      
\usepackage{xcolor}         

\usepackage{mathtools} 
\usepackage{booktabs} 
\usepackage{tikz} 

\usepackage{microtype}
\usepackage{graphicx}
\usepackage{caption}
\usepackage{subcaption}
\usepackage{booktabs} 
\usepackage{tikz-cd}
\usetikzlibrary{arrows}
\usepackage{amsmath}
\usepackage{comment}
\usepackage{amsthm}
\usepackage{amssymb}
\usepackage{url}
\usepackage{tikz}
\usepackage{tikz-qtree,tikz-qtree-compat}
\usepackage{soul}
\usepackage[ruled,vlined]{algorithm2e}

\makeatletter
\newtheorem*{rep@theorem}{\rep@title}
\newcommand{\newreptheorem}[2]{%
\newenvironment{rep#1}[1]{%
 \def\rep@title{#2 \ref{##1}}%
 \begin{rep@theorem}}%
 {\end{rep@theorem}}}
\makeatother

\newtheorem{theorem}{Theorem}
\newreptheorem{theorem}{Theorem}

\usetikzlibrary{positioning}
\newtheorem{definition}[theorem]{Definition}

\newcommand\independent{\protect\mathpalette{\protect\independenT}{\perp}}
\def\independenT#1#2{\mathrel{\rlap{$#1#2$}\mkern2mu{#1#2}}}

\title{Bounds on Causal Effects and Application to High Dimensional Data}

\author{
 Ang Li\\
  University of California, Los Angeles\\
  Computer Science Department\\
  \texttt{angli@cs.ucla.edu} \\
   \And
 Judea Pearl\\
  University of California, Los Angeles\\
  Computer Science Department\\
  \texttt{judea@cs.ucla.edu} \\
}

\begin{document}

\maketitle

\begin{abstract}
This paper addresses the problem of estimating causal effects when adjustment variables in the back-door or front-door criterion are partially observed. For such scenarios, we derive bounds on the causal effects by solving two non-linear optimization problems, and demonstrate that the bounds are sufficient. Using this optimization method, we propose a framework for dimensionality reduction that allows one to trade bias for estimation power, and demonstrate its performance using simulation studies.
\end{abstract}

\section{Introduction}\label{sec:intro}
Estimating causal effects has been encountered in many areas of industry, marketing, and health science, and it is the most critical problem in causal inference. Pearl's back-door and front-door criteria, along with the adjustment formula \cite{pearl1995causal}, are powerful tools for estimating causal effects. In this paper, the problem of estimating causal effects when adjustment variables in the back-door or front-door criterion are partially observable, or when the adjustment variables have high dimensionality, is addressed.

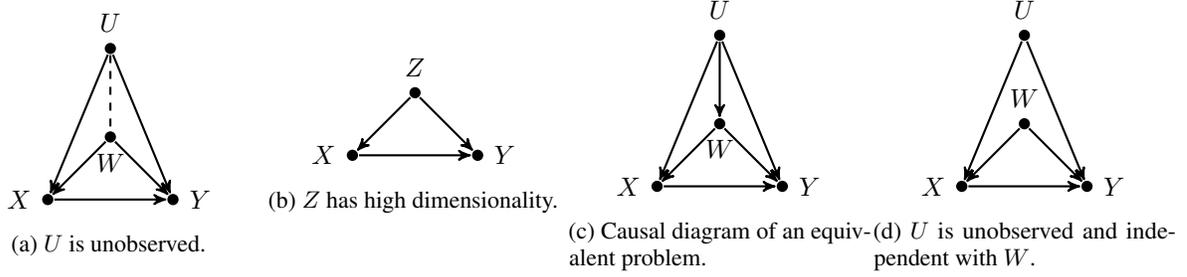
\begin{figure}
    
      \centering
        \begin{subfigure}{.24\textwidth}
            \centering
            \begin{tikzpicture}[->,>=stealth',node distance=2cm,
              thick,main node/.style={circle,fill,inner sep=1.5pt}]
              \node[main node] (1) [label=above:{$U$}]{};
              \node[main node] (2) [below =1cm of 1] [label=below:{$W$}]{};
              \node[main node] (3) [below left =1cm of 2,label=left:$X$]{};
              \node[main node] (4) [below right =1cm of 2,label=right:$Y$] {};
              \path[every node/.style={font=\sffamily\small}]
                (1) edge node {} (3)
                (1) edge node {} (4)
                (2) edge node {} (3)
                (2) edge node {} (4)
                (3) edge node {} (4);
             \draw [dashed, -] (1) -- (2);
            \end{tikzpicture}
            \caption{$U$ is unobserved.}
            \label{causal1}
        \end{subfigure}
        \begin{subfigure}{.24\textwidth}
            \centering
            \begin{tikzpicture}[->,>=stealth',node distance=2cm,
              thick,main node/.style={circle,fill,inner sep=1.5pt}]
              \node[main node] (1) [label=above:{$Z$}]{};
              \node[main node] (3) [below left =1cm of 1,label=left:$X$]{};
              \node[main node] (4) [below right =1cm of 1,label=right:$Y$] {};
              \path[every node/.style={font=\sffamily\small}]
                (1) edge node {} (3)
                (1) edge node {} (4)
                (3) edge node {} (4);
            \end{tikzpicture}
            \caption{$Z$ has high dimensionality.}
            \label{causal2}
        \end{subfigure}
        \begin{subfigure}{.24\textwidth}
            \centering
            \begin{tikzpicture}[->,>=stealth',node distance=2cm,
              thick,main node/.style={circle,fill,inner sep=1.5pt}]
              \node[main node] (1) [label=above:{$U$}]{};
              \node[main node] (2) [below =1cm of 1] [label=below:{$W$}]{};
              \node[main node] (3) [below left =1cm of 2,label=left:$X$]{};
              \node[main node] (4) [below right =1cm of 2,label=right:$Y$] {};
              \path[every node/.style={font=\sffamily\small}]
                (1) edge node {} (3)
                (1) edge node {} (2)
                (1) edge node {} (4)
                (2) edge node {} (3)
                (2) edge node {} (4)
                (3) edge node {} (4);
            \end{tikzpicture}
            \caption{Causal diagram of an equivalent problem.}
            \label{causal5}
        \end{subfigure}
        \begin{subfigure}{.24\textwidth}
            \centering
            \begin{tikzpicture}[->,>=stealth',node distance=2cm,
              thick,main node/.style={circle,fill,inner sep=1.5pt}]
              \node[main node] (1) [label=above:{$U$}]{};
              \node[main node] (2) [below =1cm of 1] [label=above:{$W$}]{};
              \node[main node] (3) [below left =1cm of 2,label=left:$X$]{};
              \node[main node] (4) [below right =1cm of 2,label=right:$Y$] {};
              \path[every node/.style={font=\sffamily\small}]
                (1) edge node {} (3)
                (1) edge node {} (4)
                (2) edge node {} (3)
                (2) edge node {} (4)
                (3) edge node {} (4);
            \end{tikzpicture}
            \caption{$U$ is unobserved and independent with $W$.}
            \label{causal6}
        \end{subfigure}
        \caption{Needed the causal effects of $X$ on $Y$.}
\end{figure}

Consider the problem of estimating the causal effects of $X$ on $Y$ when a sufficient set $W\cup U$ of confounders is partially observable (see Figure \ref{causal1}). Because $W\cup U$ is assumed to be sufficient, the causal effects are identified from measurements on $X,Y,W,$ and $U$ and can be written as
\begin{eqnarray*}
P(y|do(x)) &=& \sum_{w,u} P(y|x,w,u)P(w,u)=\sum_{w,u} \frac{P(x,y,w,u)P(w,u)}{P(x,w,u)}.
\end{eqnarray*}
However, if $U$ is unobserved, $d$-separation tells us immediately that adjusting for $W$ is inadequate by leaving the back-door path $X\xleftarrow{}U\xrightarrow{}Y$ unblocked. Therefore, regardless of sample size, the causal effects of $X$ on $Y$ cannot be estimated without bias. However, it turns out that when given a prior distribution $P(U)$, we can obtain bounds on the causal effects. We will demonstrate later that the midpoints of the bounds are sufficient for estimating the causal effects.

Bounding has been proven to be useful in causal inference. \cite{balke1997bounds} provided bounds on causal effects with imperfect compliance, \cite{tian2000probabilities} proposed bounds on     probabilities of causation, \cite{cai:etal08-r335} provided bounds on causal effects with the presence of confounded intermediate variables, and \cite{li2019unit} proposed bounds on the benefit function of a unit selection problem.

Although $P(U)$ is assumed to be given, it is usually known regardless of the model itself (e.g., $U$ stands for gender, gene type, blood type, or age). Alternatively, if costs permit, one can estimate $P(U)$ by re-testing within a small sampled sub-population.

A second problem considered in this paper is that of estimating causal effects when a sufficient set $Z$ of confounders is fully observable (see Figure \ref{causal2}), but with a high dimensionality (e.g., $Z$ has $1024$ instantiates). In such a case, a prohibitively large sample size would be required, which is generally recognized to be impractical. We propose a new framework that transforms the problem associated with Figure \ref{causal2} into an equivalent problem associated with Figure \ref{causal5} containing $W$ and $U$, which have much smaller dimensionalities (e.g., $W$ and $U$ have $32$ instantiates). We then estimate bounds on causal effects of the equivalent problem and take the midpoints as the effect estimates. We demonstrate through a simulation that this method can deliver good estimates of causal effects of the original problem.

\section{Preliminaries \& Related Works}
\label{related work}
In this section, we review the back-door and front-door criteria and their associated adjustment formulas \cite{pearl1995causal}. We use the causal diagrams in \cite{pearl1995causal, spirtes2000causation, pearl2009causality, koller2009probabilistic}. 

One key concept of a causal diagram is called $d$-separation \cite{pearl2014probabilistic}.

\begin{definition}[$d$-separation]
In a causal diagram $G$, a path $p$ is blocked by a set of nodes $Z$ if and only if 
\begin{enumerate}
    \item $p$ contains a chain of nodes $A\xrightarrow{}B\xrightarrow{}C$ or a fork $A\xleftarrow{}B\xrightarrow{}C$ such that the middle node $B$ is in $Z$ (i.e., $B$ is conditioned on), or
    \item $p$ contains a collider $A\xrightarrow{}B\xleftarrow{}C$ such that the collision node $B$ is not in $Z$, and no descendant of $B$ is in $Z$.
\end{enumerate}
If $Z$ blocks every path between two nodes $X$ and $Y$, then $X$ and $Y$ are $d$-separated conditional on $Z$, and thus are independent conditional on $Z$, denoted as $X\independent Y\ |\ Z$.
\end{definition}

With the concept of $d$-separation in a causal diagram, Pearl proposed the back-door and front-door criteria as follows:
\begin{definition}[Back-Door Criterion]
Given an ordered pair of variables $(X,Y)$ in a directed acyclic graph $G$, a set of variables $Z$ satisfies the back-door criterion relative to $(X,Y)$, if no node in $Z$ is a descendant of $X$, and $Z$ blocks every path between $X$ and $Y$ that contains an arrow into $X$.
\end{definition}
If a set of variables $Z$ satisfies the back-door criterion for $X$ and $Y$, the causal effects of $X$ on $Y$ are given by the adjustment formula:
\begin{eqnarray}
P(y|do(x)) = \sum_z P(y|x,z)P(z).
\label{adj1}
\end{eqnarray}

\begin{definition}[Front-Door Criterion]
A set of variables $Z$ is said to satisfy the front-door criterion relative to an ordered pair of variables $(X,Y)$ if
\begin{itemize}
    \item $Z$ intercepts all directed paths from $X$ to $Y$;
    \item there is no back-door path from $X$ to $Z$; and
    \item all back-door paths from $Z$ to $Y$ are blocked by $X$.
\end{itemize}
\end{definition}
If a set of variables $Z$ satisfies the front-door criterion for $X$ and $Y$, and if $P(x,Z)>0$, then the causal effects of $X$ on $Y$ are given by the adjustment formula:
\begin{eqnarray}\label{adj2}
P(y|do(x)) = \sum_z P(z|x)\sum_{x'} P(y|x',z)P(x').
\end{eqnarray}

The back-door and front-door criteria are two powerful tools for estimating causal effects; however, causal effects are not identifiable if the set of adjustment variables $Z$ is not fully observable. \cite{tian2000probabilities} provided the naivest bounds for causal effects (Equation \ref{inequ1}), regardless of the causal diagram.
\begin{eqnarray}
P(x,y) \le P(y|do(x))\le 1 - P(x,y').
\label{inequ1}
\end{eqnarray}
As the first contribution of this study, we obtain narrower bounds of the causal effects by leveraging another source of knowledge, i.e., a causal diagram behind data combined with measurements of a set $W$ (observable part of $Z$) of covariates and a prior information of a set $U$ (unobservable part of $Z$), in a causal diagram in which the bounds are solutions to two non-linear optimization problems. We illustrate that the midpoints of the bounds are sufficient for estimating the causal effects.

Using this optimization method, our second contribution is the proposal of a new framework for estimating causal effects when a set of fully observable adjustment variables $Z$ has a high dimensionality without any assumption regarding the data-generating process. \cite{maathuis2009estimating}
proposed a method of estimating causal effects when the number of covariates is larger than the sample size. However, it relies on several assumptions, including the assumption that the distribution of covariates is multivariate normal. The method is limited if the distribution of covariates is unknown or does not have accuracy estimate owing to the limitation of the sample size.

\section{Bounds on Causal Effects}
In this section, we demonstrate how bounds on causal effects with partially observable back-door or front-door variables can be obtained through non-linear optimizations.

\subsection{Partially Observable Back-Door Variables}\label{sbd}
\begin{theorem}
Given a causal diagram $G$ and a distribution compatible with $G$, let $W\cup U$ be a set of variables satisfying the back-door criterion in $G$ relative to an ordered pair $(X,Y)$, where $W\cup U$ is partially observable, i.e., only probabilities $P(X,Y,W)$ and $P(U)$ are given, the causal effects of $X$ on $Y$ are then bounded as follows:
\begin{eqnarray*}
\text{LB}\le P(y|do(x)) \le \text{UB}
\end{eqnarray*}
where LB is the solution to the non-linear optimization problem in Equation \ref{opt1} and UB is the solution to the non-linear optimization problem in Equation \ref{opt2}.
\begin{eqnarray}
&&LB=\min \sum_{w,u} \frac{a_{w,u}b_{w,u}}{c_{w,u}},\label{opt1}\\
&&UB=\max \sum_{w,u} \frac{a_{w,u}b_{w,u}}{c_{w,u}},\label{opt2}\\
&&\text{where,}\nonumber\\
&&\sum_u a_{w,u} = P(x,y,w), \sum_u b_{w,u} = P(w),\sum_u c_{w,u} = P(x,w) \text{  for all } w\in W;\nonumber\\
&&\text{ and~for~all } w\in W \text{ and } u\in U,\nonumber\\
&&b_{w,u} \ge c_{w,u} \ge a_{w,u},\nonumber\\
&&\max \{0,p(x,y,w)+p(u)-1\}\le a_{w,u} \le \min \{P(x,y,w),p(u)\},\nonumber\\
&&\max \{0,p(w)+p(u)-1\}\le b_{w,u} \le \min \{P(w),p(u)\},\nonumber\\
&&\max \{0,p(x,w)+p(u)-1\}\le c_{w,u} \le \min \{P(x,w),p(u)\}.\nonumber
\end{eqnarray}
\label{tm1}
\end{theorem}
\subsection{Partially Observable Front-Door Variables}\label{sfd}
\begin{theorem}
Given a causal diagram $G$ and distribution compatible with $G$, let $W\cup U$ be a set of variables satisfying the front-door criterion in $G$ relative to an ordered pair $(X,Y)$, where $W\cup U$ is partially observable, i.e., only probabilities $P(X,Y,W)$ and $P(U)$ are given and $P(x,W,U)>0$, the causal effects of $X$ on $Y$ are then bounded as follows:
\begin{eqnarray*}
\text{LB}\le P(y|do(x)) \le \text{UB}
\end{eqnarray*}
where LB is the solution to the non-linear optimization problem in Equation \ref{opt3} and UB is the solution to the non-linear optimization problem in Equation \ref{opt4}.
\begin{eqnarray}
&&LB=\min \sum_{w,u} \frac{b_{x,w,u}}{P(x)} \sum_{x'}\frac{a_{x',w,u}P(x')}{b_{x',w,u}},\label{opt3}\\
&&UB=\max \sum_{w,u} \frac{b_{x,w,u}}{P(x)} \sum_{x'}\frac{a_{x',w,u}P(x')}{b_{x',w,u}},\label{opt4}\\
&&\text{where,}\nonumber\\
&&\sum_u a_{x,w,u} = P(x,y,w),\sum_u b_{x,w,u} = P(x,w) \text{  for all } x\in X \text{ and } w\in W;\nonumber\\
&&\text{ and~for~all } x\in X \text{,} w\in W \text{, and } u\in U,\nonumber\\
&&b_{x,w,u}\ge a_{x,w,u},\nonumber\\
&&\max \{0,p(x,y,w)+p(u)-1\} \le a_{x,w,u} \le \min \{P(x,y,w),p(u)\},\nonumber\\
&&\max \{0,p(x,w)+p(u)-1\} \le b_{x,w,u}  \le \min \{P(x,w),p(u)\}.\nonumber
\end{eqnarray}
\label{tm2}
\end{theorem}
Notably, if any observational data (e.g., $P(U)$) are unavailable in the above theorems, we can remove that term, and the rest of non-linear optimization problems still provide valid bounds for the causal effects. In general, midpoints of bounds on causal effects are effective estimates. However, the lower (upper) bounds are also informative, which can be interpreted as the minimal (maximal) causal effects. The proofs of Theorems \ref{tm1} and \ref{tm2} are provided in the appendix.
\section{Example}
Herein, we present a simulated example to demonstrate that the midpoints of the bounds on the causal effects given by Theorem \ref{tm1} are adequate for estimating the causal effects.
\subsection{Causal Effect of a Drug}
Drug manufacturers want to know the causal effect of recovery when a drug is taken. Thus, they conduct an observational study. Here, the recovery rates of $700$ patients were recorded. A total of $192$ patients chose to take the drug and $508$ patients did not. The results of the study are shown in Table \ref{obsdata}. Blood type (type O or not) is not the only confounder of taking the drug and recovery. Another confounder is age (below the age of $70$ or not). The manufacturers have no data associated with age. They only know that $85.43\%$ of people in their region are below the age of $70$.

Because both age and blood type are confounders of taking the drug and recovery, and the observational data associated with age are unobservable, the causal effect is not identifiable.

Let $X=x$ denote the event that a patient took the drug, and $X=x'$ denote the event that a patient did not take the drug. Let $Y=y$ denote the event that a patient recovered, and $Y=y'$ denote the event that a patient did not recover. Let $W=w$ represent a patient with blood type O, and $W=w'$ represent a patient without blood type O. Let $U=u$ represent a patient below the age of $70$, and $U=u'$ represent a patient above the age of $70$. The causal diagram is shown in Figure \ref{causal6}.

\begin{table}[!htb]

    \begin{minipage}{.48\linewidth}
      \centering
      \caption{Results of an observational study considering blood type.}
      \resizebox{\columnwidth}{!}{%
            \begin{tabular}{|c|c|c|}
            \hline 
            &Drug&No Drug\\
            \hline
            \begin{tabular}{c}Blood\\type O\end{tabular}&\begin{tabular}{c}$23$ out of $36$ \\recovered\\ ($63.9\%$)\end{tabular}&\begin{tabular}{c}$145$ out of $225$ \\recovered\\ ($64.4\%$)\end{tabular}\\
            \hline
            \begin{tabular}{c}Not blood\\type O\end{tabular}&\begin{tabular}{c}$135$ out of $156$ \\recovered\\ ($86.5\%$)\end{tabular}&\begin{tabular}{c}$152$ out of $283$ \\recovered\\ ($53.7\%$)\end{tabular}\\
            \hline
            Overall&\begin{tabular}{c}$158$ out of $192$ \\recovered\\ ($82.3\%$)\end{tabular}&\begin{tabular}{c}$297$ out of $508$ \\recovered\\ ($58.5\%$)\end{tabular}\\
            \hline
            \end{tabular}%
        }
            
            \label{obsdata}    
    \end{minipage}%
    \begin{minipage}{.48\linewidth}
      \centering
      \caption{Informer view of the observational data considering blood type and age.}
            \resizebox{.8\columnwidth}{!}{%
            \begin{tabular}{|c|c|c|}
            \hline 
            &Drug&No Drug\\
            \hline
            \begin{tabular}{c}Blood\\type O\\and\\Age\\below $70$\end{tabular}&\begin{tabular}{c}$3$ out of $4$ \\recovered \\($75.0\%$)\end{tabular}&\begin{tabular}{c}$141$ out of $219$ \\recovered \\($64.4\%$)\end{tabular}\\
            \hline
            \begin{tabular}{c}Blood\\type O\\and\\Age\\above $70$\end{tabular}&\begin{tabular}{c}$20$ out of $32$ \\recovered\\ ($62.5\%$)\end{tabular}&\begin{tabular}{c}$4$ out of $6$ \\recovered\\ ($66.7\%$)\end{tabular}\\
            \hline
            \begin{tabular}{c}Not blood\\type O\\and\\Age\\below $70$\end{tabular}&\begin{tabular}{c}$135$ out of $151$ \\recovered\\ ($89.4\%$)\end{tabular}&\begin{tabular}{c}$117$ out of $224$ \\recovered\\ ($52.2\%$)\end{tabular}\\
            \hline
            \begin{tabular}{c}Not blood\\type O\\and\\Age\\above $70$\end{tabular}&\begin{tabular}{c}$0$ out of $5$ \\recovered\\ ($0.0\%$)\end{tabular}&\begin{tabular}{c}$35$ out of $59$ \\recovered\\ ($59.3\%$)\end{tabular}\\
            \hline
            Overall&\begin{tabular}{c}$158$ out of $192$ \\recovered\\ ($82.3\%$)\end{tabular}&\begin{tabular}{c}$297$ out of $508$ \\recovered\\ ($58.5\%$)\end{tabular}\\
            \hline
            \end{tabular}%
        }
            
            \label{obsdataall}
    \end{minipage}
\end{table}

An option for the manufacturers could be estimating the causal effect through the Tian-Pearl bounds in Equation \ref{inequ1} and the observational data from Table \ref{obsdata}, where
\begin{eqnarray*}
P(x,y)&=&\sum_w P(y|x,w)P(x|w)P(w)=0.2257,\\
1-P(x,y')&=&1-\sum_w P(y'|x,w)P(x|w)P(w)=0.9514.
\end{eqnarray*}
Therefore, the bounds on the causal effect estimated using Equation \ref{inequ1} are $0.2257\le P(y|do(x))\le 0.9514$, where the causal information of the covariate $W$ and the prior information $P(U)$ are not used. These bounds are not sufficiently informative to conclude the actual causal effect. Although one may believe that we can use the midpoint of the bounds (i.e., $0.5886$), the gap (i.e., $0.9514-0.2257=0.7257$) between the bounds is not small; hence, this point estimate is unconvincing.

Now, considering the proposed bounds in Theorem \ref{tm1} with the observational data from Table \ref{obsdata}. $W\cup U$ satisfies the back-door criterion, and $P(X,Y,W)$ and $P(U)$ are available. We have $12$ optimal variables in each objective function, because $W$ and $U$ are binary. With the help of the ``SLSQP'' solver  \cite{kraft1988software} in the scipy package \cite{scipy}, we obtain the bounds on the causal effect, which are $0.4728\le P(y|do(x))\le 0.9514$. The lower bound actually increased significantly, and reached close to $0.5$, which can help make decisions. The midpoint is $0.7121$. Our conclusion is then that the causal effect of recovery when taking the drug is $0.7121$. We show in the following section that this estimate of the causal effect is extremely close to the actual causal effect.

\subsection{Informer View of the Causal Effect}
An informer with access to the fully observed observational data, as summarized in Table \ref{obsdataall} (Note that although it can be verified that the data in Table \ref{obsdataall} are compatible with those in Table \ref{obsdata}, we will never know these numbers in reality), would easily calculate the causal effect of recovery when taking the drug using the adjustment formula in Equation \ref{adj1} (shown in Equation \ref{act}). The error of the estimate of the causal effect using Theorem \ref{tm1} is only $(0.7518-0.7121)/0.7518\approx 5.28\%$. 
\begin{eqnarray}
P(y|do(x))&=&\sum_{z,u} P(y|x,z,u)P(z,u)=0.7518.
\label{act}
\end{eqnarray}

\subsection{Simulation Results}
\label{simulation1}
Here, we further illustrate that the midpoints of the proposed bounds on causal effects are sufficient for estimating the causal effects, and the midpoints of the proposed bounds in Theorem \ref{tm1} are better than the midpoints of the Tian-Pearl bounds in Equation \ref{inequ1} based on a random simulation.

\begin{figure}
\begin{subfigure}{.49\textwidth}
\centering
\includegraphics[width=1\textwidth]{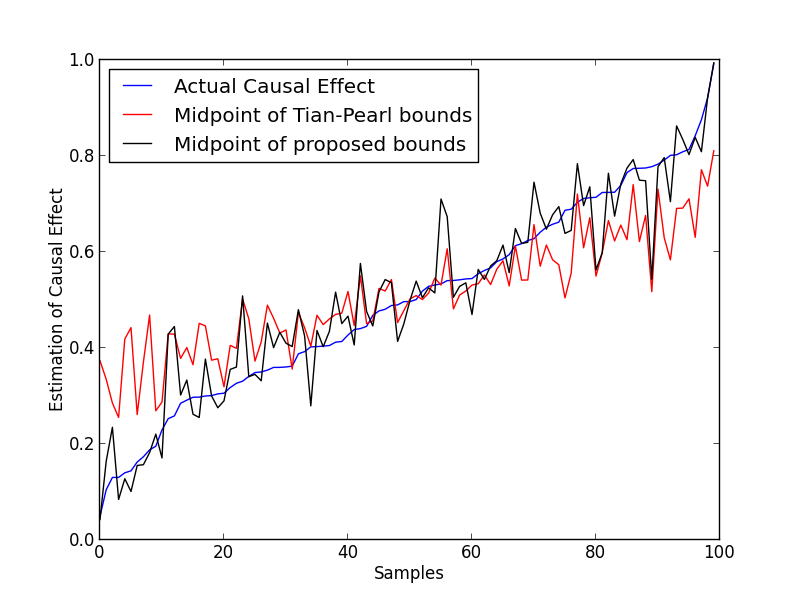}
\caption{Estimates of causal effects with partially observed confounders.}
\label{res1}
\end{subfigure}
\begin{subfigure}{.49\textwidth}
\centering
\includegraphics[width=1\textwidth]{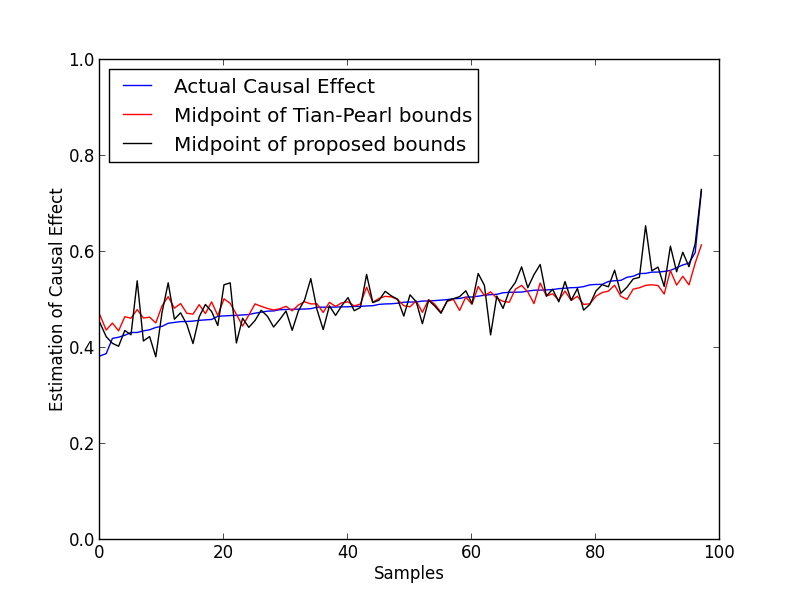}
\caption{Estimates of causal effects with high dimensional data.}
\label{res3}
\end{subfigure}
\caption{Bounds on causal effects of 100 sample distributions, where the Tian-Pearl bounds are obtained through Equation \ref{inequ1} and the proposed bounds are obtained through Theorem \ref{tm1}.}
\label{res4}
\end{figure}

We employ the simplest causal diagram in Figure \ref{causal1} with binary $W$, $U$, such that $W\cup U$ satisfies the back-door criterion. We randomly generated $1000$ sample distributions compatible with the causal diagram (the algorithm for generating the sample distributions is shown in the appendix). The average gap (upper bound $-$ lower bound) of the Tian-Pearl bounds among $1000$ samples is $0.487$, and the average gap of the proposed bounds among $1000$ samples is $0.383$. We then randomly picked $100$ out of $1000$ sample distributions to draw the graph of the actual causal effects, the midpoints of the Tian-Pearl bounds, and the midpoints of the proposed bounds. The results are shown in Figure \ref{res1}.

From Figure \ref{res1}, although both midpoints of the bounds on the causal effects are good estimates of the actual causal effects, the midpoints of the proposed bounds are much closer to the actual causal effects, particularly when the causal effects are close to $0$ and $1$. The average gap (upper bounds $-$ lower bounds), $0.383$, of the proposed bounds among $1000$ samples is much smaller than the average gap, $0.487$, of the Tian-Pearl bounds among $1000$ samples. This means that the midpoints of the proposed bounds are more convincing, because the bounds are narrower.

\section{Application to High Dimensionality of Adjustment Variables}\label{app}
Consider the problem of estimating the causal effects of $X$ on $Y$ when a sufficient set $Z$, which satisfies the back-door or front-door criterion, is fully observable (e.g., see Figure \ref{causal2}) in a causal diagram $G$ but has high dimensionality (e.g., $Z$ has $1024$ instantiates), a prohibitive large sample size would be required to estimate the causal effects, which is generally recognized to be impractical. Herein, we propose a new framework to achieve dimensionality reduction.

\subsection{Equivalent Causal Diagram with Observational Data}
\label{eqrela}

\begin{definition}[Equivalent causal diagram with observational data]
Let $G,G'$ be causal diagrams both containing nodes $X,Y$. $O$ are observational data compatible with $G$, and $O'$ are observational data compatible with $G'$. We say that $(G,O)$ is equivalent to $(G',O')$ if the causal effects of $X$ on $Y$ with $(G,O)$ is equal to the causal effects of $X$ on $Y$ with $(G',O')$.
\end{definition}

This equivalent tuple $(G',O')$ is easy to obtain. We can simply add two new nodes $W$ and $U$, and remove a node $Z$ in $G$ to obtain $G'$. Let the arrows entering $Z$ in $G$ now enter both $W$ and $U$ in $G'$, and let the arrows exiting $Z$ in $G$ now exit both $W$ and $U$ in $G'$. Finally, add an arrow from $U$ to $W$. It is easy to show that $(G,O)$ and $(G',O')$ are equivalent if the states of $Z$ are the Cartesian product of the states of $W$ and the states of $U$. Formally, we have the following theorem (the proof of the theorem is provided in the appendix),

\begin{theorem}
Let $G$ be a causal diagram containing nodes $\{V_1,...,V_{n-3},X,Y,Z\}$. Let $O$ be any observational data compatible with $G$. Suppose there exists a set of variables that satisfies the back-door or front-door criterion relative to $(X,Y)$ in $G$, then, $(G,O)$ is equivalent to $(G',O')$ ($G'$ containing nodes $\{V_1,...,V_{n-3},X,Y,W,U\}$; $O'$ are observational data compatible with $G'$), where the number of states in $W$ times the number of states in $U$ is equal to the number of states in $Z$, and the structure of $G'$ and the observational data $O'$ are obtained as follows:

Structure of $G'$:\\
Let $Parents_{G}(H)$ be the parents of $H$ in causal diagram $G$.\\
$Parents_{G'}(U)=Parents_{G}(Z)$, $Parents_{G'}(W)=Parents_{G}(Z)\cup \{U\}$,\\ $Parents_{G'}(H)=Parents_{G}(H)$ if $Z\notin Parents_{G}(H)$ for $H \in \{V_1,...,V_{n-3},X,Y\}$,\\
$Parents_{G'}(H)=Parents_{G}(H)\setminus \{Z\}\cup \{W,U\}$ \\
$~~~~~~~~~~~~~~~~~~~~~~~~~~~$if $Z\in Parents_{G}(H)$ for $H \in \{V_1,...,V_{n-3},X,Y\}$.

Note that, let $Q$ be the set of variables in $G$ that satisfies the back-door or front-door criterion relative to $(X,Y)$, then $Q'$ satisfies the back-door or front-door criterion relative to $(X,Y)$ in $G'$ , where\\
$Q'=Q$ if $Z\notin Q$,\\
$Q'=Q\setminus \{Z\}\cup \{W,U\}$ if $Z\in Q$.

Observational data:\\
Let $p$ be the number of states in $W$, and $q$ be the number of states in $U$.\\
The states of Z are the Cartesian product of the states of W and the states of U.\\
In detail, $(w_j,u_k)$ is equivalent to $z_{(j-1)*q+k}$, $w_j$ is equivalent to $\lor_{k=1}^{q}(w_j,u_k)=\lor_{k=1}^{q}z_{(j-1)*q+k}$, and $u_k$ is equivalent to $\lor_{j=1}^{p}(w_j,u_k)=\lor_{j=1}^{p}z_{(j-1)*q+k}$, i.e., $P(w_j,u_k,V)=P(z_{(j-1)*q+k},V)$ for any $V\subseteq \{V_1,...,V_{n-3},X,Y\}$.
\label{tm3}
\end{theorem}

For example, consider the causal diagram in Figure \ref{causal2} and the observational data (in the form of conditional probability tables (CPTs), where $X,Y$ are binary, and $Z$ has $4$ states.) in Table \ref{cpt1}. The causal effect, $P(y|do(x))$, through the adjustment formula in Equation \ref{adj1}, is $0.47$. Based on the construction in Theorem \ref{tm3} (see the appendix for details), we have the causal diagram in Figure \ref{causal2} with the observational data in Table \ref{cpt1} is equivalent to the causal diagram in Figure \ref{causal5} with the observational data in Table \ref{cpt2} (all nodes are binary), and we can verify that the causal effect, $P(y|do(x))$, in the causal diagram in Figure \ref{causal5} with the observational data in Table \ref{cpt2} is also $0.47$.

\begin{table}[!htb]
\begin{minipage}{.49\linewidth}
    \caption{Observational data in CPTs compatible with the causal diagram in Figure \ref{causal2}.}
    \begin{minipage}{.49\linewidth}
      \centering
        \begin{tabular}{|l|l|}
            \hline
            $P(z_1)$ & 0.3\\
            \hline
            $P(z_2)$ & 0.2\\
            \hline
            $P(z_3)$ & 0.2\\
            \hline
            $P(z_4)$ & 0.3\\
            \hline
        \end{tabular}
        \begin{tabular}{|l|l|}
            \hline
            $P(x|z_1)$ & 0.1\\
            \hline
            $P(x|z_2)$ & 0.4\\
            \hline
            $P(x|z_3)$ & 0.5\\
            \hline
            $P(x|z_4)$ & 0.7\\
            \hline
        \end{tabular}
    \end{minipage}%
    \begin{minipage}{.49\linewidth}
      \centering
        \begin{tabular}{|l|l|}
            \hline
            $P(y|x, z_1)$ & 0.2\\
            \hline
            $P(y|x', z_1)$ & 0.3\\
            \hline
            $P(y|x, z_2)$ & 0.7\\
            \hline
            $P(y|x', z_2)$ & 0.1\\
            \hline
            $P(y|x, z_3)$ & 0.6\\
            \hline
            $P(y|x',z_3)$ & 0.5\\
            \hline
            $P(y|x,z_4)$ & 0.5\\
            \hline
            $P(y|x',z_4)$ & 0.4\\
            \hline
        \end{tabular}
    \end{minipage}
    \label{cpt1}
    \end{minipage}
    \begin{minipage}{.49\linewidth}
    \caption{Observational data in CPTs compatible with the causal diagram in Figure \ref{causal5}.}
    \begin{minipage}{.49\linewidth}
      \centering
        \begin{tabular}{|l|l|}
            \hline
            $P(u)$ & 0.5\\
            \hline
        \end{tabular}
        \\
        \begin{tabular}{|l|l|}
            \hline
            $P(w|u)$ & 0.6\\
            \hline
            $P(w|u')$ & 0.4\\
            \hline
        \end{tabular}
        \begin{tabular}{|l|l|}
            \hline
            $P(x|u,w)$ & 0.1\\
            \hline
            $P(x|u,w')$ & 0.4\\
            \hline
            $P(x|u',w)$ & 0.5\\
            \hline
            $P(x|u',w')$ & 0.7\\
            \hline
        \end{tabular}
    \end{minipage}%
    \begin{minipage}{.49\linewidth}
      \centering
        \begin{tabular}{|l|l|}
            \hline
            $P(y|x, u, w)$ & 0.2\\
            \hline
            $P(y|x', u, w)$ & 0.3\\
            \hline
            $P(y|x, u, w')$ & 0.7\\
            \hline
            $P(y|x', u, w')$ & 0.1\\
            \hline
            $P(y|x, u', w)$ & 0.6\\
            \hline
            $P(y|x',u', w)$ & 0.5\\
            \hline
            $P(y|x,u', w')$ & 0.5\\
            \hline
            $P(y|x',u', w')$ & 0.4\\
            \hline
        \end{tabular}
    \end{minipage}
    
    \label{cpt2}
    \end{minipage}
\end{table}
Notably, the equivalent tuple is not unique and is transitive (i.e., if $(G,O)$ is equivalent to $(G',O')$, and $(G',O')$ is equivalent to $(G'',O'')$, then $(G,O)$ is equivalent to $(G'',O'')$). 

\subsection{Dimensionality Reduction}
Now, considering the problem in the beginning of Section \ref{app}. First, we transform the causal diagram $G$ with the compatible observational data $O$ into an equivalent tuple $(G',O')$ using Algorithm \ref{alg1} based on the construction in Theorem \ref{tm3} (note that the algorithm only construct the structure of the $G'$ and assigning the meaning of the states for $W,U$, the corresponding observatioal data $O'$ are then easy to obtain), then the new problem $(G',O')$ has the same causal effects of $X$ on $Y$ as in $(G,O)$. By picking the dimensionality of $W$ ($p$ in Algorithm \ref{alg1}), we can control the dimensionality of the new problem.

Note that, if $Z=(Z_1,Z_2,...,Z_m)$ in $G$ is a set of variables, we can repeat Algorithm \ref{alg1} for each variable in $Z$, and finally obtain $W=(W_1,W_2,...,W_m)$ and $U=(U_1,U_2,...,U_m)$, where the multiplication of the number of states in $W$ is equal to $p$. 

We then treat the new problem $(G',O')$ as a partially observable back-door or front-door variables problem in Sections \ref{sbd} and \ref{sfd}, where $P(X,Y,W)$ and $P(U)$ are given, and we can then obtain the bounds of the causal effects through Theorems \ref{tm1} and \ref{tm2}. We claim that the midpoints of the bounds are good estimates of the original causal effects. In addition, the bounds themselves will help make decisions.

\subsection{Example}\label{apps}
Consider the problem in Figure \ref{causal2}, where $X$ and $Y$ are binary and $Z$ has $256$ states. We randomly generated a distribution $P(X,Y,Z)$ that is compatible with the causal diagram. Because we know the exact distribution, we can easily obtain the causal effects through Equation \ref{adj1}. The causal effect $P(y|do(x))$ is $0.5527$ (the algorithm for generating the distribution is shown in the appendix).

Now, we transform the causal diagram with the observational data into an equivalent tuple $(G',O')$ ($G'$ is shown in Figure \ref{causal5}) using Algorithm \ref{alg1} ($p=16$). We obtain the variable $W$ of $16$ states and the variable $U$ of $16$ states in $G'$ ($(w_j,u_k)$ is equivalent to $z_{(j-1)*16+k}$). We are then forced to use only observational data $P(X,Y,W)$ and $P(U)$ (the construction of $P(X,Y,W)$ and $P(U)$ is shown in the appendix), and based on Theorem \ref{tm1}, with the ``SLSQP'' solver, we obtain the bounds on the causal effect $p(y|do(x))$, which are $0.4595\le P(y|do(x))\le 0.7012$. We see the midpoint, $0.5804$, is extremely close to the actual causal effect, $0.5527$. 

\resizebox{.9\columnwidth}{!}{%
\begin{algorithm}[H]
    \SetAlgoLined
    \SetKwData{ProbZ}{prob\_z}
    \SetKwData{ProbYXZ}{prob\_y\_xz}
    \SetKwData{LB}{lb}
    \SetKwData{LBGraph}{lb\_graph}
    \SetKwData{UBGraph}{ub\_graph}
    \SetKwData{UB}{ub}
    \SetKwData{CPT}{cpt}
    \SetKwData{N}{n}
    \SetKwFunction{UnifRand}{random-uniform}
    \SetKwFunction{PNS}{pns-bounds}
    \SetKwFunction{PNSGraph}{pns-graph}
    \SetKwFunction{AppendToResult}{append-result}
    \SetKwFunction{NumberofInsG}{num\_states\_in\_G}
    \SetKwFunction{NumberofInsGp}{num\_states\_in\_G'}
    \SetKwFunction{Alg}{generate-cpt}
    \SetKwFunction{ParentsG}{Parents\_in\_G}
    \SetKwFunction{ParentsGp}{Parents\_in\_G'}
    \SetKwInOut{Input}{input}\SetKwInOut{Output}{output}
    \Input{A $n$ nodes, $(X_1,X_2,...,X_{n-3},X,Y,Z)$, causal diagram $G$ and compatible $O$,\\$p$, the number of states in $W$ in $G'$ of the equiv. tuple $(G',O')$.\\}
    \Output{A $n+1$ nodes, $(X_1,X_2,...,X_{n-3},X,Y,W,U)$, causal diagram $G'$,\\Maping relation $M_1:\text{state of W}\xrightarrow[]{}\text{state of Z}$,\\Maping relation $M_2:\text{state of U}\xrightarrow[]{}\text{state of Z}$.}
    \Begin{
    $m$ $\leftarrow$ \NumberofInsG{$Z$}\;
    \If{$m \mod p = 0$}{
    $q$ $\leftarrow$ $m / p$\;
    }{
    \Else{$q$ $\leftarrow$ $m / p + 1$\;
    }
    }
    \tcp{Set the virtual states for $Z$ such that the probability is $0$.}

    \NumberofInsG{$Z$} $\leftarrow$ $p\times q$\; 
    
    \For{$H$ in $\{X_1,...,X_{n-3},X,Y\}$}{
        \NumberofInsGp{$H$} $\leftarrow$ \NumberofInsG{$H$}\;
        \If{$Z \in $\ParentsG{$H$}}{
            \ParentsGp{$H$} $\leftarrow$ \ParentsG{$H$}$\setminus\{Z\}\cup\{W,U\}$\;
        }{
        \Else{
            \ParentsGp{$H$} $\leftarrow$ \ParentsG{$H$}\;
        }
        }
    }
    
    \NumberofInsGp{$W$} $\leftarrow$ $p$\;
    \NumberofInsGp{$U$} $\leftarrow$ $q$\;
    \ParentsGp{$W$} $\leftarrow$ \ParentsG{$Z$}$\cup \{U\}$\;
    \ParentsGp{$U$} $\leftarrow$ \ParentsG{$Z$}\;
    \For{$i \leftarrow 1$ \KwTo $p$}{
        $M_1(w_i)$ $\leftarrow$ $\lor_{k=1}^{q}z_{(i-1)*q+k}$\;
    }
    \For{$i \leftarrow 1$ \KwTo $q$}{
        $M_2(u_i)$ $\leftarrow$ $\lor_{j=1}^{p}z_{(j-1)*q+i}$\;
    }
    
    }
    \caption{Generate Equivalent Tuple}
    \label{alg1}
\end{algorithm}%
}

Finally, lets consider how many samples are required for each method. According to \cite{roscoe1975fundamental}, each state needs at least $30$ samples, and therefore, the exact solution by Equation \ref{adj1} requires $2\times 2\times 256\times 30=30720$ samples. However, the proposed bounds based on Theorem \ref{tm1} only requires $max(2\times 2\times 16, 16)\times 30=1920$ samples. If the sample size is still unacceptable, we can use another equivalent tuple with $W$ having $8$ states and $U$ having $32$ states, we then only require $max(2\times 2\times 8, 32)\times 30=960$ samples to obtain the bounds on the causal effects.

\subsection{Simulation Results}
\label{simulation2}
Similarly to the previous simulation, we further illustrate that the bounds on the causal effects of the proposed framework are sufficient for estimating the original causal effects.

Once again, by employing the simplest causal diagram in Figure \ref{causal2}, where $X$ and $Y$ are binary and $Z$ has $256$ states. We randomly generated $100$ sample distributions compatible with the causal diagram (the algorithm for generating the distributions are shown in the appendix). The average gap (upper bound $-$ lower bound) of the Tian-Pearl bounds among $100$ samples is $0.5102$, and the average gap of the proposed bounds through Theorems \ref{tm3} and \ref{tm1} among $100$ samples is $0.0676$. We then draw the graph of the actual causal effects, the midpoints of the Tian-Pearl bounds, and the midpoints of the proposed bounds through Theorems \ref{tm3} and \ref{tm1}. The results are shown in Figure \ref{res3}.

From Figure \ref{res3}, both midpoints of the bounds on the causal effects are good estimates of the actual causal effects, whereas the midpoints of the proposed bounds are slightly closer to the actual causal effects, particularly when the causal effects are close to $0$ and $1$. Although the trend of the Tian-Pearl bounds is also close to the actual causal effects, the Tian-Pearl bounds are more likely to be parallel with the x-axis. Here, the Tian-Pearl bounds perform well because, in high-dimensionality cases, the randomly generated distributions are more likely to yield causal effects of approximately $0.5$. However, the average gap of the proposed bounds among $100$ samples, $0.0676$, is much smaller than the average gap of the Tian-Pearl bounds among $100$ samples, $0.5102$. This means that the midpoints of the proposed bounds are more convincing, because the bounds are narrower.

\section{Discussion}
\label{discussion}
Here, we discuss additional features of bounds on causal effects. 

First, if a whole set of back-door or front-door variables are unobserved, the causal effects have the naivest bounds in Equation \ref{inequ1}. When the back-door or front-door variables are gradually observed, the bounds of the causal effects become increasingly narrow. Finally, when the back-door or front-door variables are fully observed, the bounds shrink into point estimates, which are identifiable. This also tells us that, when we pick $p$ in Algorithm \ref{alg1}, we should pick the largest $p$ for which the sample size is sufficient to estimate the observational distributions. 

Next, bounds in Theorems \ref{tm1} and \ref{tm2} are given by non-linear optimizations. Therefore, the quality of the bounds also depends on the optimization solver. The examples and simulated results in this paper are all obtained from the simplest ``SLSQP'' solver from 1988. The quality of the bounds can be improved if more advanced solvers are applied. Inspired by the idea of Balke's linear programming \cite{balke1997probabilistic}, we may obtain parametric solutions to non-linear optimizations in Theorems \ref{tm1} and \ref{tm2}, we then do not need a non-linear optimization solver. However, the problem related to a non-linear optimization solver is not the scope of this paper.

In addition, the constraints in Theorems \ref{tm1} and \ref{tm2} are only based on the basic back-door or front-door criterion. We can also add constraints of independencies in a specific graph. For instance, $W$ and $U$ are independent in the causal diagram of Figure \ref{causal6}, we can then add the constraints that reflect $P(W)$ and $P(U)$ as being independent. The greater the number of constraints that are added to the optimizations, the better the bounds we can obtain.

Moreover, if one believes they have a sufficient sample size to estimate causal effects with high dimensionality adjustment variables, the framework in Section \ref{app} could be evidence validating whether the sample size is indeed sufficient.

Next, in Section \ref{app}, we transformed $(G,O)$ into $(G',O')$ to obtain the bounds on causal effects with high dimensionality adjustment variables. However, for a tuple $(G,O)$, multiple equivalent tuples exist by picking a different $p$ in Algorithm \ref{alg1}, and each of the equivalent tuple has bounds for the original causal effects. We can compute bounds for as many equivalent tuples as we want and take the maximal lower bounds and the minimal upper bounds.

Finally, based on numerous experiments, we realized that when $P(U)$ or $P(W)$ is specific (i.e., closer to $0$ or $1$), the proposed bounds are almost identified (i.e., the bounds shrink to point estimates). Therefore, in practice, we can always pick the equivalent tuple to transform, in which the $P(U)$ or $P(W)$ is close to $0$ or $1$.

\section{Conclusion}
\label{conclusion}
We demonstrated how to estimate causal effects when adjustment variables in the back-door or front-door criterion are partially observable by bounding the causal effects using solutions to non-linear optimizations. We provided examples and simulated results illustrating that the proposed method is sufficient to estimate the causal effects. We also proposed a framework for estimating causal effects when the adjustment variables have a high dimensionality. In summary, we analyzed and demonstrated how causal effects can be gained in practice using a causal diagram.

\bibliography{arxiv01}

\newpage

\appendix

\section{Proof of Theorem \ref{tm1}}
\begin{reptheorem}{tm1}
Given a causal diagram $G$ and a distribution compatible with $G$, let $W\cup U$ be a set of variables satisfying the back-door criterion in $G$ relative to an ordered pair $(X,Y)$, where $W\cup U$ is partially observable, i.e., only probabilities $P(X,Y,W)$ and $P(U)$ are given, the causal effects of $X$ on $Y$ are then bounded as follows:
\begin{eqnarray*}
\text{LB}\le P(y|do(x)) \le \text{UB}
\end{eqnarray*}
where LB is the solution to the non-linear optimization problem in Equation \ref{opt771} and UB is the solution to the non-linear optimization problem in Equation \ref{opt772}.
\begin{eqnarray}
&&LB=\min \sum_{w,u} \frac{a_{w,u}b_{w,u}}{c_{w,u}},\label{opt771}\\
&&UB=\max \sum_{w,u} \frac{a_{w,u}b_{w,u}}{c_{w,u}},\label{opt772}\\
&&\text{where,}\nonumber\\
&&\sum_u a_{w,u} = P(x,y,w), \sum_u b_{w,u} = P(w),\sum_u c_{w,u} = P(x,w) \text{  for all } w\in W;\nonumber\\
&&\text{ and~for~all } w\in W \text{ and } u\in U,\nonumber\\
&&b_{w,u} \ge c_{w,u} \ge a_{w,u},\nonumber\\
&&\max \{0,p(x,y,w)+p(u)-1\}\le a_{w,u} \le \min \{P(x,y,w),p(u)\},\nonumber\\
&&\max \{0,p(w)+p(u)-1\}\le b_{w,u} \le \min \{P(w),p(u)\},\nonumber\\
&&\max \{0,p(x,w)+p(u)-1\}\le c_{w,u} \le \min \{P(x,w),p(u)\}.\nonumber
\end{eqnarray}
\begin{proof}
To show that the LB and UB bound the actual causal effects, we only need to show that there exists a point in feasible space of the non-linear optimization that $\sum_{w,u} \frac{a_{w,u}b_{w,u}}{c_{w,u}}$ is equal to the actual causal effects.\\
Since $W\cup U$ satisfies the back-door criterion, by adjustment formula in Equation \ref{adj1}, we have,\\
\begin{eqnarray*}
P(y|do(x)) & =& \sum_{w,u} P(y|x,w,u)P(w,u)\\
&=&\sum_{w,u} \frac{P(x,y,w,u)P(w,u)}{P(x,w,u)}
\end{eqnarray*}
Let
\begin{eqnarray*}
&&a_{w,u}=P(x,y,w,u)\\
&&b_{w,u}=P(w,u)\\
&&c_{w,u}=P(x,w,u)
\end{eqnarray*}
We now show that the above set of $a_{w,u},b_{w,u},c_{w,u}$ are in feasible space.\\
We have,
\begin{eqnarray*}
\text{  for } w&\in& W\\
\sum_u a_{w,u} &=& \sum_u P(x,y,w,u)= P(x,y,w)\\
\sum_u b_{w,u} &=& \sum_u P(w,u)=P(w)\\
\sum_u c_{w,u} &=& \sum_u P(x,w,u)=P(x,w)
\end{eqnarray*}
and,
\begin{eqnarray*}
\text{ ~~for~~all } &w\in W& \text{ and~~ } u\in U\nonumber\\
b_{w,u}= P(w,u) &\ge& P(x,w,u) = c_{w,u}\nonumber\\
c_{w,u} = P(x,w,u) &\ge& P(x,y,w,u) = a_{w,u}\nonumber\\
a_{w,u} =P(x,y,w,u)&\le& \min \{P(x,y,w),p(u)\}\nonumber\\
b_{w,u}= P(w,u) &\le& \min \{P(w),p(u)\}\nonumber\\
c_{w,u}= P(x,w,u) &\le& \min \{P(x,w),p(u)\}\nonumber\\
a_{w,u}=P(x,y,w,u) &\ge& \max \{0,p(x,y,w)+p(u)-1\}\nonumber\\
b_{w,u}= P(w,u) &\ge& \max \{0,p(w)+p(u)-1\}\nonumber\\
c_{w,u}= P(x,w,u) &\ge& \max \{0,p(x,w)+p(u)-1\}\nonumber
\end{eqnarray*}
Therefore, the above set of $a_{w,u},b_{w,u},c_{w,u}$ are in feasible space, and thus, the UB and LB bound the actual causal effects.
\end{proof}
\end{reptheorem}

\section{Proof of Theorem \ref{tm2}}
\begin{reptheorem}{tm2}
Given a causal diagram $G$ and distribution compatible with $G$, let $W\cup U$ be a set of variables satisfying the front-door criterion in $G$ relative to an ordered pair $(X,Y)$, where $W\cup U$ is partially observable, i.e., only probabilities $P(X,Y,W)$ and $P(U)$ are given and $P(x,W,U)>0$, the causal effects of $X$ on $Y$ are then bounded as follows:
\begin{eqnarray*}
\text{LB}\le P(y|do(x)) \le \text{UB}
\end{eqnarray*}
where LB is the solution to the non-linear optimization problem in Equation \ref{opt773} and UB is the solution to the non-linear optimization problem in Equation \ref{opt774}.
\begin{eqnarray}
&&LB=\min \sum_{w,u} \frac{b_{x,w,u}}{P(x)} \sum_{x'}\frac{a_{x',w,u}P(x')}{b_{x',w,u}},\label{opt773}\\
&&UB=\max \sum_{w,u} \frac{b_{x,w,u}}{P(x)} \sum_{x'}\frac{a_{x',w,u}P(x')}{b_{x',w,u}},\label{opt774}\\
&&\text{where,}\nonumber\\
&&\sum_u a_{x,w,u} = P(x,y,w),\sum_u b_{x,w,u} = P(x,w) \text{  for all } x\in X \text{ and } w\in W;\nonumber\\
&&\text{ and~for~all } x\in X \text{,} w\in W \text{, and } u\in U,\nonumber\\
&&b_{x,w,u}\ge a_{x,w,u},\nonumber\\
&&\max \{0,p(x,y,w)+p(u)-1\} \le a_{x,w,u} \le \min \{P(x,y,w),p(u)\},\nonumber\\
&&\max \{0,p(x,w)+p(u)-1\} \le b_{x,w,u}  \le \min \{P(x,w),p(u)\}.\nonumber
\end{eqnarray}
\begin{proof}
To show that the LB and UB bound the actual causal effects, we only need to show that there exists a point in feasible space of the non-linear optimization that $\sum_{w,u} \frac{b_{x,w,u}}{P(x)} \sum_{x'}\frac{a_{x',w,u}P(x')}{b_{x',w,u}}$ is equal to the actual causal effects.\\
Since $W\cup U$ satisfies front-door criterion and $P(u,W,U)>0$, by adjustment formula in Equation \ref{adj2}, we have,\\
\begin{eqnarray*}
P(y|do(x)) & =& \sum_{w,u} P(w,u|x)\sum_{x'} P(y|x',w,u)P(x')\\
&=& \sum_{w,u} \frac{P(x,w,u)}{P(x)} \sum_{x'}\frac{P(x',y,w,u)P(x')}{P(x',w,u)}
\end{eqnarray*}
Let
\begin{eqnarray*}
&&a_{x,w,u}=P(x,y,w,u)\\
&&b_{x,w,u}=P(x,w,u)
\end{eqnarray*}
Similarly to the proof of Theorem \ref{tm1}, it is easy to show that the above set of $a_{x,w,u}, b_{x,w,u}$ are in feasible space, and therefore, LB and UB bound the actual causal effects.
\end{proof}
\end{reptheorem}

\section{Proof of Theorem \ref{tm3}}

\begin{reptheorem}{tm3}
Let $G$ be a causal diagram containing nodes $\{V_1,...,V_{n-3},X,Y,Z\}$. Let $O$ be any observational data compatible with $G$. Suppose there exists a set of variables that satisfies the back-door or front-door criterion relative to $(X,Y)$ in $G$, then, $(G,O)$ is equivalent to $(G',O')$ ($G'$ containing nodes $\{V_1,...,V_{n-3},X,Y,W,U\}$; $O'$ is observational data compatible with $G'$), where the number of states in $W$ times the number of states in $U$ is equal to the number of states in $Z$, and the structure of $G'$ and the observational data $O'$ are obtained as follows:

Structure of $G'$:\\
Let $Parents_{G}(H)$ be the parents of $H$ in causal diagram $G$.\\
$Parents_{G'}(U)=Parents_{G}(Z)$, $Parents_{G'}(W)=Parents_{G}(Z)\cup \{U\}$,\\ $Parents_{G'}(H)=Parents_{G}(H)$ if $Z\notin Parents_{G}(H)$ for $H \in \{V_1,...,V_{n-3},X,Y\}$,\\
$Parents_{G'}(H)=Parents_{G}(H)\setminus \{Z\}\cup \{W,U\}$ \\
$~~~~~~~~~~~~~~~~~~~~~~~~~~~$if $Z\in Parents_{G}(H)$ for $H \in \{V_1,...,V_{n-3},X,Y\}$.

Note that, let $Q$ be the set of variables in $G$ that satisfies the back-door or front-door criterion relative to $(X,Y)$, then $Q'$ satisfies the back-door or front-door criterion relative to $(X,Y)$ in $G'$ , where\\
$Q'=Q$ if $Z\notin Q$,\\
$Q'=Q\setminus \{Z\}\cup \{W,U\}$ if $Z\in Q$.

Observational data:\\
Let the number of states in $W$ be $p$, and let the number of states in $U$ be $q$.\\
The states of Z is the Cartesian product of the states of W and the states of U.\\
In detail, $(w_j,u_k)$ is equivalent to $z_{(j-1)*q+k}$, $w_j$ is equivalent to $\lor_{k=1}^{q}(w_j,u_k)=\lor_{k=1}^{q}z_{(j-1)*q+k}$, and $u_k$ is equivalent to $\lor_{j=1}^{p}(w_j,u_k)=\lor_{j=1}^{p}z_{(j-1)*q+k}$, i.e., $P(w_j,u_k,V)=P(z_{(j-1)*q+k},V)$ for any $V\subseteq \{V_1,...,V_{n-3},X,Y\}$.
\begin{proof}
First, we show that $Q'$ satisfies the back-door or front-door criterion relative to $(X,Y)$ in $G'$.\\

If $Q$ satisfies the back-door criterion relative to $(X,Y)$ in $G$, we need to show that, 
\begin{itemize}
    \item no node in $Q'$ is a descendant of $X$.
    \item $Q'$ blocks every path between $X$ and $Y$ that contains an arrow into $X$.
\end{itemize}
It is easy to show that if there is a node in $Q'$ that is a descendant of $X$ in $G'$, then there is a node in $Q$ that is a descendant of $X$ in $G$. And if there is a path between $X$ and $Y$ that contains an arrow into $X$ does not blocked by $Q'$ in $G'$, then there is a path between $X$ and $Y$ that contains an arrow into $X$ does not blocked by $Q$ in $G$. Thus, $Q'$ satisfies the back-door criterion relative to $(X,Y)$ in $G'$.
Similarly, we can show that if $Q$ satisfies the front-door criterion relative to $(X,Y)$ in $G$, then $Q'$ satisfies the front-door criterion relative to $(X,Y)$ in $G'$.

Now, we show that $(G,O)$ is equivalent to $(G',O')$, i.e., show that $P(y|do(x))$ is the same between $(G,O)$ and $(G',O')$. Suppose $Q$ satisfies the back-door criterion relative to $(X,Y)$ in $G$. By adjustment formula in Equation \ref{adj1}, we have,\\
$P(y|do(x))=\sum_{q\in Q}P(y|x,q)\times P(q)=\sum_{q\in Q}\frac{P(x,y,q)\times P(q)}{P(x,q)}$.\\
And in $G'$, \\
$P(y|do(x))=\sum_{q\in Q'}P(y|x,q)\times P(q)=\sum_{q\in Q'}\frac{P(x,y,q)\times P(q)}{P(x,q)}$,\\
it is obviously that these two causal effects are the same, because $P(w_j,u_k,V)=P(z_{(j-1)*q+k},V)$ for any $V\subseteq \{V_1,...,V_{n-3},X,Y\}$.\\
Similarly, we can show that if $Q$ satisfies the front-door criterion relative to $(X,Y)$ in $G$, $(G,O)$ is equivalent to $(G',O')$.
\end{proof}
\end{reptheorem}

\section{Simulation Algorithm for Generating Sample Distributions in Sections \ref{simulation1}, \ref{apps}, and \ref{simulation2}}

\begin{algorithm}[H]
    \SetAlgoLined
    \SetKwData{ProbZ}{prob\_z}
    \SetKwData{ProbYXZ}{prob\_y\_xz}
    \SetKwData{LB}{lb}
    \SetKwData{LBGraph}{lb\_graph}
    \SetKwData{UBGraph}{ub\_graph}
    \SetKwData{UB}{ub}
    \SetKwData{S}{s}
    \SetKwData{P}{p}
    \SetKwData{Sum}{sum}
    \SetKwFunction{UnifRand}{uniform-random}
    \SetKwFunction{PNS}{pns-bounds}
    \SetKwFunction{PNSGraph}{pns-graph}
    \SetKwFunction{AppendToResult}{append-result}
    \SetKwFunction{Sample}{sample}
    \SetKwFunction{NumberofIns}{num-instantiates}
    \SetKwInOut{Input}{input}\SetKwInOut{Output}{output}
    \Input{$n$ causal diagram nodes ($X_1,...,X_n$)\\Distribution $D$}
    \Output{$n$ conditional probability tables for $P(X_i|Parents(X_i))$}
    \Begin{
    \For{$i \leftarrow 1$ \KwTo $n$}{
        \S $\leftarrow$ \NumberofIns{$X_i$}
        
        \P $\leftarrow$ \NumberofIns{$Parents(X_i)$}
        
        \For{$k \leftarrow 1$ \KwTo $p$}{
            \Sum $\leftarrow$ $0$
            
                \For{$j \leftarrow 1$ \KwTo $s$}{
                    $a_j$ $\leftarrow$ \Sample{$D$}
                    
                    \Sum $\leftarrow$ \Sum $+$ $a_j$
                }
            
                \For{$j \leftarrow 1$ \KwTo $s$}{
                    $P(x_{i_j}|Parents(X_i)_k)$ $\leftarrow$ $a_j/$\Sum

                }
        }
    }
    }
    \caption{Generate-cpt()}
    \label{alg222}
\end{algorithm}

In our simulation studies, we set $D$ in Algorithm \ref{alg222} to the uniform distribution.

\section{Construction of the Data in Table \ref{cpt2} of Section \ref{eqrela}}
\begin{eqnarray*}
P(u,w)&=&P(z_1),\\
P(u,w')&=&P(z_2),\\
P(u',w)&=&P(z_3),\\
P(u',w')&=&P(z_4),\\
P(u)&=&P(u,w)+P(u,w')=P(z_1)+P(z_2)=0.5,\\
P(w|u)&=&P(u,w)/p(u)=P(z_1)/P(u)=0.3/0.5=0.6,\\
P(w|u')&=&P(u',w)/p(u')=P(z_3)/(1-P(u))=0.2/0.5=0.4,\\
P(x|u,w)&=&P(x|z_1)=0.1,\\
P(x|u,w')&=&P(x|z_2)=0.4,\\
P(x|u',w)&=&P(x|z_3)=0.5,\\
P(x|u',w')&=&P(x|z_4)=0.7,\\
P(y|x,u,w)&=&P(y|x,z_1)=0.2,\\
P(y|x',u,w)&=&P(y|x',z_1)=0.3,\\
P(y|x,u,w')&=&P(y|x,z_2)=0.7,\\
P(y|x',u,w')&=&P(y|x',z_2)=0.1,\\
P(y|x,u',w)&=&P(y|x,z_3)=0.6,\\
P(y|x',u',w)&=&P(y|x',z_3)=0.5,\\
P(y|x,u',w')&=&P(y|x,z_4)=0.5,\\
P(y|x',u',w')&=&P(y|x',z_4)=0.4.
\end{eqnarray*}

\section{Construction of the Distribution in Section \ref{apps}}
Instead of providing the resulting 1024 rows of the observational data, we provide the details for regenerating the observational data as following steps. 
\begin{itemize}
    \item Generate $P(X,Y,Z)$ using Algorithm \ref{alg222}.
    \item Let $P(X,Y,w_j,u_k)=P(X,Y,z_{(j-1)*16+k})$.
    \item Let $P(X,Y,w_j)=\sum_{k=1}^{q}P(X,Y,w_j,u_k).$
    \item Let $P(X,Y,u_k)=\sum_{j=1}^{p}P(X,Y,w_j,u_k).$
    \item Let $P(u_k)=\sum_{X,Y}P(X,Y,u_k).$
\end{itemize}
For example, 
\begin{eqnarray*}
P(u_1) &=& \sum_{X,Y}P(X,Y,u_1)\\
&=&P(x,y,u_1)+P(x,y',u_1)+P(x',y,u_1)+P(x',y',u_1)\\
&=&\sum_{j=1}^{16}P(x,y,w_j,u_1)+\sum_{j=1}^{16}P(x,y',w_j,u_1)+\sum_{j=1}^{16}P(x',y,w_j,u_1)+\sum_{j=1}^{16}P(x',y',w_j,u_1)\\
&=&\sum_{j=1}^{16}P(x,y,z_{(j-1)*16+1})+\sum_{j=1}^{16}P(x,y',z_{(j-1)*16+1})+\\
&&\sum_{j=1}^{16}P(x',y,z_{(j-1)*16+1})+\sum_{j=1}^{16}P(x',y',z_{(j-1)*16+1}),\\
P(x,y,w_1) &=& \sum_{k=1}^{16}P(x,y,w_1,u_k)\\
&=&\sum_{k=1}^{16}P(x,y,z_{k}).
\end{eqnarray*}


\end{document}